\documentclass[twocolumn, 9pt]{extarticle}
\usepackage{amsmath,amsthm,amssymb,nicefrac}
\usepackage{mathtools}
\usepackage{lineno,hyperref}
\usepackage[table,x11names,dvipsnames,table]{xcolor}
\usepackage{authblk}
\usepackage{subcaption,booktabs,siunitx}
\usepackage{graphicx}
\usepackage{multirow}
\usepackage[nolist,nohyperlinks]{acronym}
\usepackage[numbers,sort&compress]{natbib}
\usepackage{tabularx}
\usepackage{titlesec}
\usepackage{url}
\usepackage{xurl}
\usepackage{amsfonts}       
\usepackage{nicefrac}       
\usepackage{microtype}      
\usepackage{lipsum}
\usepackage{algorithm}
\usepackage{algorithmic}
\usepackage{multicol}
\usepackage{array}
\usepackage{makecell} 
\usepackage{adjustbox}
\usepackage{wrapfig}
\usepackage[acronym]{glossaries}
\usepackage{float}
\usepackage[hang,flushmargin]{footmisc} 
\usepackage[group-separator={,},output-decimal-marker={.}]{siunitx}
\usepackage{geometry}
\makeglossaries
\newacronym{ANNs}{ANNs}{Artificial Neural Networks}
\newacronym{MLP}{MLP}{Multilayer Perceptron}
\newacronym{FC}{FC}{Fully Connected}
\newacronym{ML}{ML}{Machine Learning}
\newacronym{FL}{FL}{Federated Learning}
\newacronym{PFL}{PFL}{Personalised Federated Learning}
\newacronym{MTL}{MTL}{Multitask Learning}
\newacronym{ReLU}{ReLU}{Rectified Linear Unit}
\newacronym{PLWD}{PLWD}{People Living with Dementia}
\newacronym{UTI}{UTI}{Urinary Tract Infection}
\newacronym{WCSS}{WCSS}{Within-Cluster Sum of Squares}
\newacronym{LR}{LR}{Logistic Regression}
\newacronym{PIR}{PIR}{Passive Infra-Red}
\newacronym{t-SNE}{t-SNE}{t-distributed Stochastic Neighbour Embedding}

\titlespacing*{\section}
{0pt}{2ex plus 1ex minus 1ex}{2ex plus 1ex}
\titlespacing*{\subsection}
{0pt}{.5ex plus .25ex minus .25ex}{.5ex plus .25ex}

\makeatletter
\renewcommand{\citet}[1]{%
  \def\natexlab##1{}%
  \citeauthor{#1} (\citeyear{#1})~[\citenum{#1}]%
}

\makeatother

\DeclareCaptionLabelFormat{bold}{\textbf{(#2)}}
\graphicspath{{Figures/}}  

\makeatletter
\def\@maketitle{
\raggedright
\newpage
  \noindent
  \vspace{0cm}
  \let \footnote \thanks
    {\hskip -0.4em \huge \textbf{{\@title}} \par}
    \vskip 1.5em
    {\large
      \lineskip .5em
      \begin{tabular}[t]{l}
      \raggedright
        \@author
      \end{tabular}\par}
    \vskip 1em
  \par
  \vskip 1.5em
  }
\makeatother

\begin{document}

\title{Urinary Tract Infection Detection in Digital Remote Monitoring: Strategies for Managing Participant-Specific Prediction Complexity}

\author[1, 2]{Kexin Fan\thanks{kexin.fan23@imperial.ac.uk}}
\author[1, 2]{Alexander Capstick\thanks{alexander.capstick19@imperial.ac.uk}}
\author[1, 2, 3, 4]{Ramin Nilforooshan}
\author[1, 2, 5]{Payam Barnaghi}

\affil[1]{Imperial College London}
\affil[2]{Care Research and Technology Centre, UK Dementia Research Institute}
\affil[3]{University of Surrey}
\affil[4]{Surrey and Borders Partnership NHS Trust}
\affil[5]{Data Research, Innovation and Virtual Environments (DRIVE) Unit, The Great Ormond Street Hospital}

\setcounter{Maxaffil}{0}
\renewcommand\Affilfont{\itshape\small}

\date{}  
\maketitle

\begin{abstract}
Urinary tract infections (UTIs) are a significant health concern, particularly for people living with dementia (PLWD), as they can lead to severe complications if not detected and treated early.
This study builds on previous work that utilised machine learning (ML) to detect UTIs in PLWD by analysing in-home activity and physiological data collected through low-cost, passive sensors. The current research focuses on improving the performance of previous models, particularly by refining the Multilayer Perceptron (MLP), to better handle variations in home environments and improve sex fairness in predictions by making use of concepts from multitask learning.
This study implemented three primary model designs: feature clustering, loss-dependent clustering, and participant ID embedding which were compared against a baseline MLP model. The results demonstrated that the loss-dependent MLP achieved the most significant improvements, increasing validation precision from 48.92\% to 72.60\% and sensitivity from 27.44\% to 70.52\%, while also enhancing model fairness across sexes.
These findings suggest that the refined models offer a more reliable and equitable approach to early UTI detection in PLWD, addressing participant-specific data variations and enabling clinicians to detect and screen for UTI risks more effectively, thereby facilitating earlier and more accurate treatment decisions.
The code used in this study is available at the following link: \url{https://github.com/Kexin-Fan/Multi-Source-Analysing.git}.
\end{abstract}

\section{Introduction}

A \gls{UTI} is an infection affecting the urinary system, involving either the lower urinary tract alone or both the lower and upper tracts \cite{uti_definition}. The gold standard for \gls{UTI} diagnosis involves detecting pathogens in a urine culture alongside clinical symptoms. However, solely depending on clinical signs like dysuria, increased frequency, and suprapubic pain results in a 33\% error rate \cite{uti_diagnosis}. While urine culture improves prediction accuracy, it requires 24-48 hours, creating a trade-off between speed and precision. Moreover, the satisfactory sensitivity and specificity achieved in the laboratory often cannot be replicated in primary care settings \cite{uti_dipslide}.
Dementia, a comorbidity associated with symptomatic \glspl{UTI}, increases the risk of bowel and bladder incontinence, complicating diagnosis in this population \cite{uti_older_adults}. This challenge is intensified as \gls{PLWD} often struggle to report symptoms accurately, leading to potential delays in diagnosis until infections require hospitalisation \cite{uti_hospitalisation, uti_plwd_diagnosis_difficulty}. Therefore, improved targeted testing could provide more effective care.

To address the challenges of clinical \gls{UTI} diagnosis in \gls{PLWD}, which can be time-consuming, resource-intensive, and prone to inaccuracies, two key approaches—\gls{ML} and low-cost monitoring devices—have been implemented. These methods enable early, automated detection of \glspl{UTI} without relying on symptom reporting by \gls{PLWD} and before obvious symptoms arise, while minimising excessive testing. Although several studies have explored \gls{UTI} risk prediction in younger adults, these findings often do not generalise to older populations \cite{uti_ml2}. One study used in-home \gls{PIR} sensors to detect \glspl{UTI} in older adults but did not incorporate \gls{ML} techniques \cite{uti_sensor}. Another study combined in-home sensors with unsupervised \gls{ML}; however, the performance was suboptimal, relying on self-recorded physiological measurements and thus lacking full automation \cite{uti_ml1}.

The Minder platform is one of the most advanced systems integrating \gls{ML} with digital remote monitoring. Developed in collaboration with clinicians and user groups, it is designed for the remote monitoring and management of \gls{PLWD}, using low-cost sensors in their homes to continuously and automatically collect environmental and physiological data. The platform is active, with the dataset being continuously updated. The final dataset also contains collected information, including labelled \gls{UTI} data \cite{minder_platform, palermo2023tihm}. Using a \gls{UTI} symptom-specific subdataset, a previous study applied supervised \gls{ML} on Minder platform data for early \gls{UTI} detection \cite{capstick2024digital_UTI_prediction}. Several models, including \gls{LR}, XGBoost, and \gls{MLP}, were tested, with \gls{LR} achieving the best performance, showing 65.3\% sensitivity and 70.9\% specificity in predicting \glspl{UTI} in unseen participants. Although not deeply explored, the study noted that the models generally performed better for male participants than female. Additionally, variations in natural physiology, dementia type, stage, and environmental factors like home layout, caregiver presence, and sensor performance across different \gls{PLWD} homes may have significantly impacted data representation and the accuracy of \gls{UTI} label predictions.

Building on the previous \gls{ML} work \cite{capstick2024digital_UTI_prediction}, this study focuses on refining the \gls{MLP} model, which provides greater flexibility for structural and input adjustments compared to XGBoost and \gls{LR}, in order to better address home variations and improve sex fairness. The key contributions of this research are: (1) \gls{ML}: The study proposes methods to handle data from diverse sources, enhancing the model's adaptability; (2) \gls{UTI} Prediction: The study maintains the accuracy of previous predictions while ensuring the model can accommodate data variations and represent all households equitably, achieving sex fairness. These improvements can be integrated into the ongoing Minder study algorithm to enhance early detection.

\section{Background and Related Work}

\subsection{ML Approaches}

Having data points coming from sources of variable reliability has inspired \gls{FL}, a \gls{ML} approach allowing decentralised devices or institutions to collaboratively train a model without sharing their data. Developed primarily to address data privacy issues, this method involves participants training the model locally with their own data and sending only model updates (e.g., gradients or weights) to a central server \cite{kairouz2021_FL}. 
Hospital-based studies have shown that \gls{FL} models are effective for disease prediction, but they also revealed that while some predictive factors of disorders are shared, many are specific to individual institutions \cite{oh2018_two_hospitals}. \gls{FL} often encounters challenges with non-identically distributed data among clients, leading to difficulties in model convergence and consistency. Variations in clients' computational resources and network conditions can cause inefficiencies, and as more clients join, managing and coordinating them becomes increasingly complex, affecting scalability \cite{kairouz2021_FL}.

Consequently, to address the major challenge of heterogeneity in training data from different sources, \gls{PFL} is employed. According to \cite{tan2022_PFL}, there are two major \gls{PFL} strategies: Global Model Personalisation and Learning Personalised Models. Global Model Personalisation trains a global model using standard \gls{FL} and then personalises it for each client through additional local training. In contrast, Learning Personalised Models creates individual models for each client by modifying the \gls{FL} aggregation process, with methods like \gls{MTL}, which seeks to jointly learn multiple related tasks, allowing knowledge from one task to benefit others and improve overall generalisation performance. \gls{MTL} uses methods like sharing important features between tasks, grouping similar tasks together, finding relationships between tasks, and breaking down complex tasks into simpler parts to enhance performance \cite{zhang2021survey_multitask_learning}. Given that the study involves non-uniform data sources, it can be considered a ``feature homogeneous multi-task environment" \cite{alex_sources}. Therefore, this study employs \gls{MTL} methods, specifically focusing on sharing key features across tasks and grouping similar tasks.

In \gls{ML}, cross-entropy is a widely used loss criterion for classification tasks that quantifies the difference between true class labels and predicted probabilities, prompting the model to adjust its outputs toward more accurate predictions \cite{cross_entropy}. During training, the cross-entropy reduction loss curve often reveals an "elbow" point, where the rate of loss reduction slows significantly. The presence of an elbow suggests that the model is approaching convergence, indicating it has learned most patterns in the training data and is nearing optimal performance with the current data and hyperparameters. The elbow marks a point where additional training epochs yield only marginal improvements. Therefore, the losses at the elbow can indicate the difficulty of prediction.

\subsection{The Minder Dataset}

Building on the methodology from previous research on \gls{UTI} prediction using the Minder dataset, this study uses the same subdataset and feature engineering techniques \cite{capstick2024digital_UTI_prediction}. As a longitudinal study, we have access to further data, with the training set covering 28 June 2021, to 1 January 2023, and the testing set from 1 January 2023, to 1 January 2024. Initially, the training set included 69 patients, and the test set had 55 but 4 patients were removed from the test set due to incomplete demographic information, leaving 51 patients. There are 41 participants common to both the train and test sets. Therefore the 10 patients unique to the test set were either excluded or retained based on the method used.

In the original work, patient data was first sorted by date. Data gaps exceeding one day between records triggered segmentation into smaller continuous segments, with segments shorter than three days discarded. Within qualifying segments, \gls{UTI} labels were confirmed for one day and then extended to cover three days on either side of that day, effectively converting one verified label into seven days of \gls{UTI} labels. Hence, the final training dataset comprised 1,839 data points, while the test dataset contained 1,145 data points.

The 20 input features used in this study are adopted from previous research and are all symptom-specific \cite{capstick2024digital_UTI_prediction}, comprising both raw and engineered components. The raw features include: (1) activation frequency in the bathroom, bedroom, hallway, kitchen, and lounge; (2) mean and standard deviation of nocturnal heart and respiratory rates; and (3) nocturnal awake occurrences. The engineered features include: (4) frequency of bathroom use during day and night, with moving averages and percentage changes; (5) mean and standard deviation of time taken to move from any location to the bathroom; (6) daily entropy in \gls{PIR} sensor activations; and (7) the number of previous \gls{UTI}s recorded to date.

\section{Methods}

Proposed methods are summarised into two main approaches. The first approach, based on \gls{MTL} principles, predicts labels by grouping participants into artificially defined clusters—either by loss values at the elbow epoch (loss clusters) or by feature pattern similarity (feature clusters). This assumes that different homes share patterns that can enhance mutual learning, and grouping makes the identification of these patterns more efficient. Two model designs are presented: the ``Fully Separated \gls{MLP}," which operates as five independent models with separate paths for each cluster, and the ``Final Layer Separated \gls{MLP}," which shares two common hidden layers but uses distinct output layers ("heads") for each cluster. These designs parallel Learning Personalised Models and Global Model Personalisation in \gls{PFL} \cite{tan2022_PFL}. The second approach predicts labels using participant information by incorporating embeddings. Embeddings transform sparse participant or source data into continuous features integrated into the model \cite{dahouda2021deep_embedding}.

The experiments were conducted in two stages: hyperparameter tuning, and training and testing. During hyperparameter tuning, a grid search was conducted to optimise learning rates (0.001, 0.005, 0.01) and dropout rates (0.000, 0.200, 0.500). The model's structure was determined based on prior \gls{ML} experience and is detailed below. Cross-validation was used, with 80\% of the training data randomly selected for training and 20\% for validation. Optimal hyperparameters were identified by the lowest average validation loss at the final epoch across all ten folds (Table \ref{tb:minder_model_structure}). The best model was then trained five times using a resampling strategy, with 80\% of the training data randomly selected each time to prevent overfitting, and then tested on the entire test dataset. Notably, during the resampling process, stratification by participants was ensured by sampling 80\% of each participant's data points, rather than 80\% of the entire dataset, to maintain the representativeness of each participant in the training process.

\begin{table}[b]
    \centering
    \caption{Minder Models' Hyperparametres}
    \setlength{\tabcolsep}{1.5pt} 
    \footnotesize
    \begin{tabular}{c c c}
        \toprule
        \textbf{MLP} & \makecell{\textbf{Learning} \\ \textbf{Rate}} & \makecell{\textbf{Dropout} \\ \textbf{Rate}} \\
        \midrule
        \textbf{Baseline} & 0.001 & 0.500 \\
        \midrule
        \textbf{Two Feature Clusters} & 0.001 & 0.000 \\
        \textbf{Four Feature Clusters} & 0.001 & 0.500 \\
        \midrule
        \textbf{Fully Separated Loss Dependent} & 0.001 & 0.000 \\
        \textbf{Final Layer Separated Loss Dependent} & 0.001 & 0.000 \\
        \midrule
        \textbf{ID Embedding} & 0.001 & 0.000 \\
        \bottomrule
    \end{tabular}
    \label{tb:minder_model_structure}
\end{table}

Notably, for the Minder dataset, data was collected over an extended period, and so the train/test dataset split was based on date to align with real-world use cases. 
This resulted in the train/validation split being performed using cross-validation, while the train/test split was based on a temporal division, providing a comprehensive assessment of the methods’ performance under both conditions.

\paragraph{Baseline}

A baseline \gls{MLP} model was created to serve as a reference point for comparison with other more advanced models. This model features two hidden layers of sizes 30 and 10, using \gls{ReLU} activation, with the Adam optimiser \cite{kingma2014adam} and cross-entropy loss. All other models developed are based on structural modifications of the baseline \gls{MLP}.

\paragraph{Feature Clustering}

K-means clustering was applied to the 20 input features of the training dataset, which included 69 participants. Silhouette scores were calculated to determine the optimal number of clusters, measuring an object's similarity to its own cluster versus others, with higher values indicating better-defined clusters \cite{silhouette}. The scores showed that two clusters had the highest score, followed by four clusters. Consequently, models based on two clusters (Two Feature Clusters \gls{MLP}) and four clusters (Four Feature Clusters \gls{MLP}) were tested. The remaining 10 participants in the test dataset, with no overlap with the training dataset, were predicted into these clusters using the K-means clustering algorithm. The models employed a Final Layer Separated structure, diverging at the final output layer.

\paragraph{Loss Dependent}

\begin{figure}[t]
    \centering
    \begin{subfigure}[b]{\linewidth}
        \centering
        \includegraphics[width=0.7\linewidth]{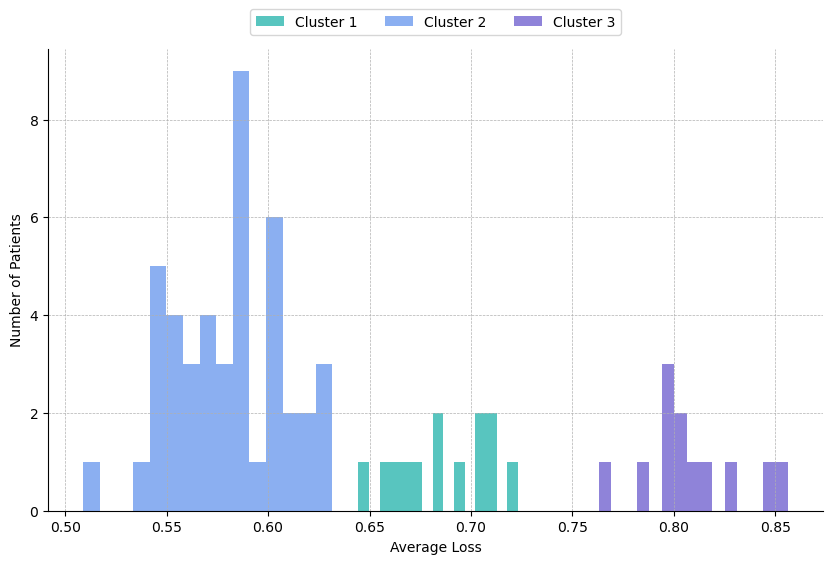}
        \caption{}
        \label{fig:minder_train_clusters}
    \end{subfigure}
    
    \begin{subfigure}[b]{\linewidth}
        \centering
        \includegraphics[width=0.7\linewidth]{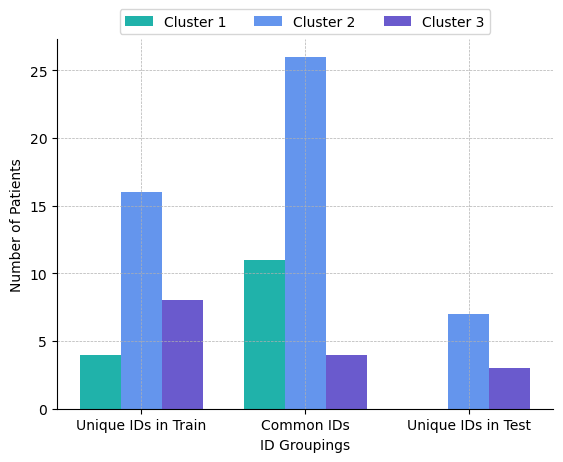}
        \caption{}
        \label{fig:minder_cluster}
    \end{subfigure}
    \caption{Loss Analysis and Clustering at Elbow Epoch: \textbf{\subref{fig:minder_train_clusters}:} Clusters of average losses for participants from the training dataset at epoch 3, identified as the 'elbow' epoch; \textbf{\subref{fig:minder_cluster}:} Clusters of average losses for participants by ID groupings: training on unique train IDs and common IDs, predicting unique test IDs.}
    \label{fig:minder_clusters}
\end{figure}

Data was first trained on the Baseline \gls{MLP} model to obtain loss information, with the training loss curve revealing an elbow point at the third epoch. Consequently, the losses at the third epoch for each \gls{PLWD} were averaged and displayed in a histogram (Figure \ref{fig:minder_train_clusters}). K-means clustering was applied to group these average losses, with the optimal number of clusters determined using the elbow method. This method plots the \gls{WCSS} against different cluster numbers to find the point where the \gls{WCSS} decrease slows, indicating the best balance between compactness and cluster number \cite{k-means_clustering}. The analysis suggested that three clusters were optimal. Each participant in the training dataset was then assigned a loss cluster label.

The Baseline \gls{MLP} parameters, including weights and gradients, were recorded at the third epoch. The test participants were then processed through the model to record their losses, which were then averaged, and clusters were predicted using the K-means algorithm. The final cluster assignments are shown in Figure \ref{fig:minder_cluster}. Both Fully Separated and Final Layer Separated architectures were employed for the Loss Dependent approach.

\paragraph{ID Embedding}

The ID Embedding \gls{MLP} incorporated an embedding layer using participants' IDs as embeddings. The 10 participants present in the test data but absent from the training data were excluded, as their new IDs would lack a pre-trained embedding vector, resulting in incompatibility with the model's established structure. The embedding dimension was set to 8, and the generated vectors were concatenated with the input data before being processed by the model.

\section{Results and Interpretation}

The cross-validation results of the models with the selected hyperparameters, along with the resampling training and testing results, are presented in Table \ref{tb:validation_minder} and Table \ref{tb:test_minder} respectively. The train/test metrics generally exceeded the train/validation metrics, suggesting that individual \gls{PLWD} exhibit highly distinct patterns. Participants' variability makes training on past data points more informative than on data points from different participants at the same time. This is evidenced by the large standard deviations in cross-validation results, likely due to the high heterogeneity among participants.

Consistent with previous findings, male participants consistently exhibit higher performance metrics (Precision, Sensitivity, and Accuracy) across all models compared to female participants. The occurrence of overall performance metrics sometimes being higher than both female and male metrics is likely due to the fact that if a division by zero occurs during calculation—such as when there are no positive predictions or true positive cases—the result is set to 0. Consequently, during multiple fold cross-validation and repeated resampling training, if only one sex has no positive predictions or true positive cases, it can significantly lower the metrics for that sex, thereby artificially inflating the overall metrics.

\begin{table*}[ht]
    \centering
    \caption{Validation Metrics (\%) for Different MLP Models Across Two Sexes}
    \scalebox{1.0}{
    \resizebox{\textwidth}{!}{
    \setlength{\tabcolsep}{1pt}
    \begin{tabular}{c c c c c c c c c c c}
        \toprule
        \multirow{2}{*}{\textbf{MLP}} & \multicolumn{3}{c}{\textbf{Precision}} & \multicolumn{3}{c}{\textbf{Sensitivity}} & \multicolumn{3}{c}{\textbf{Accuracy}} \\
        \cmidrule(lr){2-4} \cmidrule(lr){5-7} \cmidrule(lr){8-10}
        & \textbf{Female} & \textbf{Male} & \textbf{Overall} & \textbf{Female} & \textbf{Male} & \textbf{Overall} & \textbf{Female} & \textbf{Male} & \textbf{Overall} \\
        \midrule
        \textbf{Baseline} & 17.45(28.07) & 50.03(44.88) & 48.92(34.84) & 4.95(9.99) & 30.67(39.28) & 27.44(31.47) & 52.27(27.97) & 83.61(13.92) & 70.42(12.14) \\
        \midrule
        \textbf{Two Feature Clusters} & 22.20(39.44) & 53.63(40.84) & 51.17(31.56) & 3.13(6.92) & 40.97(39.44) & 36.24(33.44) & 43.99(28.14) & 86.13(11.12) & 69.60(12.13) \\
        \textbf{Four Feature Clusters} & 7.37(17.60) & 43.67(45.04) & 44.57(43.61) & 3.66(8.08) & 39.00(42.94) & 34.51(39.28) & 54.95(35.02) & 85.95(12.51) & 73.79(17.05) \\
        \midrule
        \textbf{Fully Separated Loss Dependent} & 35.31(37.82) & 67.84(39.66) & 71.00(20.43) & 31.71(39.17) & 57.80(39.10) & 66.55(29.30) & 61.52(28.51) & 87.92(9.98) & 78.41(10.08) \\
        \textbf{Final Layer Separated Loss Dependent} & 48.73(41.86) & 64.91(38.84) & 72.60(18.55) & 39.65(41.48) & 60.00(41.62) & 70.52(30.09) & 75.10(17.53) & 86.68(12.00) & 79.61(10.34) \\
        \midrule
        \textbf{ID Embedding} & 23.07(35.18) & 50.82(40.55) & 49.77(30.97) & 14.52(24.59) & 47.36(42.57) & 49.59(36.74) & 44.22(26.03) & 85.15(14.18) & 68.74(13.60) \\
        \bottomrule
    \end{tabular}
    }
    }
    \caption*{Values in parentheses represent standard deviations.}
    \label{tb:validation_minder}
\end{table*}

\begin{table*}[ht]
    \centering
    \caption{Test Metrics (\%) for Different MLP Models Across Two Sexes}
    \scalebox{1.0}{
    \resizebox{\textwidth}{!}{
    \setlength{\tabcolsep}{1pt}
    \begin{tabular}{c c c c c c c c c c c}
        \toprule
        \multirow{2}{*}{\textbf{MLP}} & \multicolumn{3}{c}{\textbf{Precision}} & \multicolumn{3}{c}{\textbf{Sensitivity}} & \multicolumn{3}{c}{\textbf{Accuracy}} \\
        \cmidrule(lr){2-4} \cmidrule(lr){5-7} \cmidrule(lr){8-10}
        & \textbf{Female} & \textbf{Male} & \textbf{Overall} & \textbf{Female} & \textbf{Male} & \textbf{Overall} & \textbf{Female} & \textbf{Male} & \textbf{Overall} \\
        \midrule
        \textbf{Baseline} & 51.84(11.44) & 84.36(2.31) & 76.15(2.31) & 51.49(1.94) & 79.74(5.55) & 64.57(3.36) & 68.89(1.97) & 90.50(1.38) & 82.01(1.06) \\
        \midrule
        \textbf{Two Feature Clusters} & 42.71(1.10) & 70.13(5.04) & 58.56(3.23) & 26.96(7.14) & 70.43(7.52) & 46.96(4.38) & 67.16(1.12) & 90.13(1.68) & 81.10(1.07) \\
        \textbf{Four Feature Clusters} & 36.14(6.36) & 62.76(15.62) & 50.78(10.68) & 16.89(7.03) & 42.26(16.81) & 28.56(11.49) & 66.93(0.63) & 86.94(2.83) & 79.07(1.58) \\
        \midrule
        \textbf{Fully Separated Loss Dependent} & 38.03(2.55) & 69.00(2.74) & 53.03(3.59) & 32.00(6.84) & 62.61(11.24) & 46.08(4.56) & 63.78(2.11) & 89.24(1.54) & 79.23(1.19) \\
        \textbf{Final Layer Separated Loss Dependent} & 38.12(7.59) & 68.18(2.05) & 50.24(5.66) & 35.56(6.16) & 51.30(10.27) & 42.80(5.34) & 62.84(4.98) & 88.03(1.22) & 78.13(2.02) \\
        \midrule
        \textbf{ID Embedding} & 31.79(9.77) & 77.88(3.71) & 67.32(7.02) & 6.45(2.47) & 57.39(5.82) & 31.27(3.75) & 66.05(2.52) & 87.67(1.36) & 78.74(1.58) \\
        \bottomrule
    \end{tabular}
    }
    }
    \caption*{Values in parentheses represent standard deviations.}
    \label{tb:test_minder}
\end{table*}

\paragraph{Baseline}

For the Baseline \gls{MLP}, performance metrics showed a holistic improvement from validation to testing. The test metrics were generally satisfactory, with an overall precision of 76.15\% (±2.31\%), overall sensitivity of 64.57\% (±3.36\%), and an overall accuracy of 82.01\% (±1.06\%), despite the persistence of a notable sex bias. 
This improvement likely results from the difference between the train/validation and train/test splits—where the former used cross-validation across participants, while the latter relied on temporal data from the same participants—indicating that learning from a participant’s previous data provides more valuable insights than cross-sectional data from other participants.

\paragraph{Feature Clustering}

The Two Feature Clusters \gls{MLP} generally showed improved performance during cross-validation but performed worse during testing compared to the Simple \gls{MLP}. Clustering by features essentially creates artificial groups, where data points with similar features are assumed to come from the same group. The improvement in the validation set demonstrated the feasibility of this method. However, during testing, these generated clusters may have overshadowed the individual patterns of each participant, possibly due to shifts in their data over time. These shifts could have led to data points being incorrectly assigned to clusters using the previously defined K-means algorithm, resulting in reduced performance.

\paragraph{Loss Dependent}

During the cross-validation process, the two Loss Dependent \gls{MLP}s generally outperformed all other models. The Final Layer Separated \gls{MLP} achieved overall precision, sensitivity, and accuracy of 72.60\% (±18.55\%), 70.52\% (±30.09\%), and 79.61\% (±10.34\%) respectively, slightly outperforming the Fully Separated \gls{MLP}, which achieved 71.00\% (±20.43\%), 66.55\% (±29.30\%), and 78.41\% (±10.08\%). These two models also showed the smallest standard deviations among all models during cross-validation, indicating they effectively capture shared hidden patterns across different \gls{PLWD}, demonstrating both consistency and generalisability. Notably, these models demonstrated improved fairness between female and male participants in the validation metrics, with a higher proportion of female participants showing significant gains from specific loss cluster paths.

\begin{figure}[h]
\centerline{\includegraphics[width=0.7\linewidth]{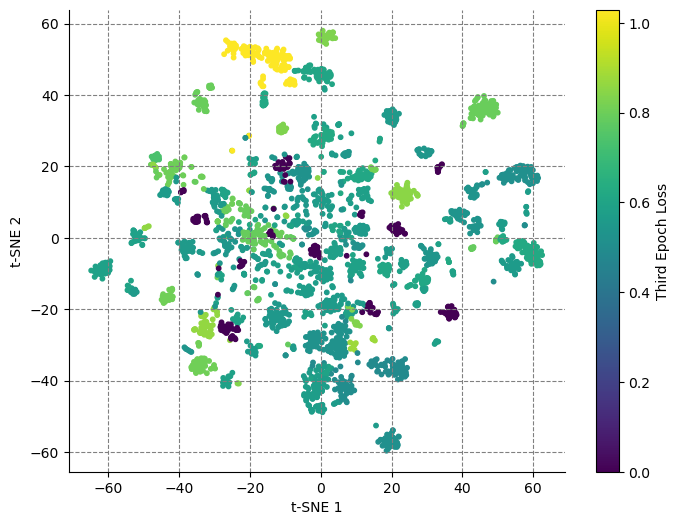}}
\caption{A \gls{t-SNE} visualisation of data points from both the training and test datasets. The colour gradient indicates the corresponding loss values of each data point at the third epoch.}
\label{fig:tsne}
\end{figure}

A \gls{t-SNE} plot provides insights into why the proposed framework performs well (Figure \ref{fig:tsne}). \gls{t-SNE} is a non-linear dimensionality reduction technique that visualises high-dimensional data in 2D or 3D while preserving local relationships, making it effective for revealing hidden clusters or patterns \cite{t-sne}. In the plot, the 2D \gls{t-SNE} embeddings of both training and test data at the third epoch (elbow epoch) show dark purple points with lower losses, scattered but forming small, distinct clusters. These clusters likely correspond to data from individual participants, indicating consistent patterns within each participant's data. More notable is the dense cluster of yellow points, which indicates higher loss values; these points are prominently grouped together, implying that participants with higher losses share specific underlying patterns.

However, during testing, neither of these models displayed significantly better results compared to other models. The clustering method used is vulnerable to shifts in patterns among new participants, leading to misclassification into incorrect loss clusters which highlights a potential limitation of the Loss Dependent approach in accommodating new participants with different data patterns.

\paragraph{ID Embedding}

The ID Embedding method did not show improvement during cross-validation or training. This outcome is not unexpected, considering the relatively small dataset size. We hypothesise that each participant had an insufficient number of data points for the model to effectively learn the patterns.

\section{Discussion and Conclusion}

The current study builds on prior \gls{ML} work identifying \gls{UTI} risk in \gls{PLWD} using activity and physiological data from low-cost, passive sensors \cite{capstick2024digital_UTI_prediction}. The goal was to enhance early detection by refining the \gls{MLP} model to account for home variations and improve sex fairness. To achieve this, three model designs were implemented: feature clustering, loss-dependent clustering, and participant ID embedding. The Feature Clustering and Loss Dependent \glspl{MLP} adapted well to varying home environments, despite potential challenges with new participants. Specifically, the Loss Dependent models notably improved sex fairness in the validation metrics.

Building on prior \gls{ML} practices, this study aligned its models with \gls{PFL} approaches, structuring them based on either the Learning Personalised Models or Global Model Personalisation methods, and incorporating \gls{MTL} techniques such as feature-sharing across tasks and grouping similar tasks \cite{tan2022_PFL}. Within a feature-homogeneous multitask framework \cite{alex_sources}, the study aimed to strike a balance between creating personalised, specific representations and identifying shared patterns to enhance performance, in line with \gls{MTL} goals. This balance was examined through decisions on using uniform models, separate pathways within models, or multiple models, and was further demonstrated by applying the elbow method to training loss reduction curves and the \gls{WCSS} curve for determining optimal cluster numbers.

The current models did not fully address the significant sex differences in prediction performance, likely due to the inherently more chaotic feature patterns in females. \gls{UTI} is one of the most pronounced infectious diseases with sex differences, as adult premenopausal women are 40 times more likely to experience \glspl{UTI} than adult men \cite{uti_sex_difference}. This increased susceptibility in females is influenced by anatomical, hormonal, and immune factors, such as a shorter urethra and fluctuating estrogen levels \cite{ingersoll2017sex_uti_why_different}. Additionally, sex may be associated with other demographics like age, dementia duration, and carer status, which were not explored in this study, adding complexity to predictions. Despite these challenges, Loss Dependent models showed better alignment in precision and sensitivity between sexes compared to other models, indicating greater robustness.

This study has several limitations worth acknowledging. First, the number of clusters was manually determined through visual inspection rather than automated methods, potentially introducing observer bias. The model is also susceptible to inductive bias, where assumptions made by the learning algorithm to generalise from training data can lead to misclassification. This is particularly concerning when the model assumes the number and structure of clusters based on training data, making it vulnerable to shifts in data patterns, especially with the introduction of new participants with different characteristics \cite{kairouz2021_FL}. Additionally, ID embedding may be less effective for participants with limited data, potentially reducing model performance for certain subsets. Future research could explore integrating semi-supervised learning to address this. Finally, although the study identified different groups through clustering, the Feature Clustering and Loss Dependent \gls{MLP}s evenly assigned clusters across models. Future work should refine this approach by enabling models to differentiate between clusters, identify the least reliable ones, and apply additional processing to those data points.

In conclusion, this study demonstrates the effectiveness of using loss information to group participants and integrating specialised pathways in \gls{ML} models in complex, heterogeneous settings. These methods not only provide valuable theoretical insights in the \gls{MTL} domain but also hold promise for enhancing clinical decision-making. They have the potential to be integrated into the ongoing Minder platform, potentially improving patient outcomes and resource use.

\section*{Code and Data Availability}

The current study was conducted using Python 3.11.5, with libraries including NumPy (1.26.4), Pandas (2.2.2), Scikit-Learn (1.4.2), and PyTorch (2.1.2). The Minder dataset can be obtained from the author and affiliated institution upon reasonable request. The code used in this study is available at the following link: \url{https://github.com/Kexin-Fan/Multi-Source-Analysing.git}.

\section*{Acknowledgement}
This study is funded by the UK Dementia Research Institute (UK DRI) Care Research and Technology Centre funded by the Medical Research Council (MRC)(grant number: UKDRI-7002), and the UKRI Engineering and Physical Sciences Research Council (EPSRC) PROTECT Project (grant number: EP/W031892/1). Infrastructure support for this research was provided by the NIHR Imperial Biomedical Research Centre (BRC) and the UKRI Medical Research Council (MRC). The funders were not involved in the study design, data collection, data analysis or writing the manuscript. We also would like to thank the UK DRI Care Research and Technology Centre's team and acknowledge their contribution in conducting the Minder study and sharing the data for this research.

\end{document}